\documentclass{aa}

\usepackage{graphicx} 
\usepackage{txfonts} 
\usepackage[latin1]{inputenc}
\usepackage{longtable} 

\begin{document} 

   \authorrunning{Poncelet et al.}

   \title{A new analysis of the nucleus of NGC 1068 \\ with MIDI observations}

   \author{A.  Poncelet \inst{1,2},\ G.  Perrin \inst{2}\ \and
   H. Sol\inst{1}\ }

   \institute{LUTH, Observatoire de Paris, 92195 Meudon Cedex\ \and LESIA,
 Observatoire de Paris, 92195 Meudon Cedex\\ \email{anne.poncelet@obspm.fr} }

   \date{Received , 2005; accepted , 2005}

   \abstract{We present a new analysis of the very first mid-infrared N-band
   long-baseline interferometric observations of an extragalactic source: the
   nucleus of the Seyfert 2 galaxy NGC 1068, obtained with MIDI (Mid-InfrareD
   Interferometer), the mid-infrared beamcombiner at the European Southern
   Observatory (ESO) Very Large Telescope Interferometer (VLTI).  The
   resolution of $\lambda$/B = 10 mas allows to study the compact central
   core of the galaxy between 8 and 13 $\mu$m.  Both visibility measurements
   and MIDI spectrum are well reproduced with a simple radiative transfer model
   with two concentric spherical components. The derived angular sizes and
   temperatures are about 35 and 83 mas, and 361 K and 226 K for these
   two components respectively.  Some other evidences strongly support such low
   temperatures. This modeling also provides the variation of optical depth as
   a function of wavelength for the extended component across the N-band
   pointing towards the presence of amorphous silicate grains. This shows that
   MIDI actually carried out the first direct observations of the distribution of dust around the central
   engine. Together with other observational pieces, we are able to move a step
   forward in the reconstruction of the picture drawn for AGNs.

   \keywords{galaxies: NGC 1068 -- galaxies: Seyfert -- galaxies: nuclei --
radiative transfer -- infrared: galaxies -- \\ techniques: interferometic}}

   \maketitle

%________________________________________________________________
\section{Introduction} 
%________________________________________________________________

NGC 1068 is one of the brightest and nearest Seyfert 2 galaxies. Located at a
distance of 14.4 Mpc, it is unique for the study of the active galactic nucleus
(AGN) it harbours. The nucleus of NGC 1068 has been well studied in the entire
spectrum, from X-rays and UV to radio wavelengths, including the optical, near-
and mid-infrared wavelength ranges. The location of all the components emitting
at different wavelengths has been revisited in Galliano et al.
(\cite{Galliano}).  NGC 1068 was classified as a Seyfert 2 with respect to the
narrow emission lines it emits.  However, Antonucci \& Miller
(\cite{Antonucci}) have discovered broad, polarized emission lines, which
suggest the presence of a Seyfert 1 nucleus, hidden by a geometrically and
optically thick dusty torus surrounding it. Thus, they laid the foundation
stone of the unified scheme of AGN. Several theoretical studies of AGN tori
were reported by Pier \& Krolik (\cite{PierKrolik1,PierKrolik2}), Granato \&
Danese (\cite{Granato}), Efstathiou et al.  (\cite{Efstathiou}).  All these
models consider radiative transfer in an anisotropic, but homogenous dusty
torus. Nenkova et al. (\cite{Nenkova}) considered radiative transfer between
several clouds of dust along radial rays through the torus, and overcame some
difficulties of the former homogeneous models for the description of the
spectral energy distributions. More recently, Schartmann et al.
(\cite{Schartmann}) describe the three dimensional treatment of transfer
radiative through dusty tori at hydrostatic equilibrium and succeed in
reproducing both the mean large aperture spectra from UV to far-IR of several
Seyfert type 1 galaxies, and the recent MIDI observations of two specified
Seyfert type 2: Circinus and NGC 1068, except for the 9.7~$\mu$m silicate
feature. Geometrical and dynamical properties of dusty tori are
theoretically investigated, among others, by Krolik \& Begelman (\cite{K&B}),
Zier \& Biermann (\cite{Zier}), and  Beckert \& Duschl (\cite{Beckert}).

An alternative explanation for the unified scheme of AGNs is given by
Elvis (\cite{Elvis}), who actually proposes a simple and empirical scenario of
funnel-shaped thin shell, warm and highly ionised outflow that could
account for all the structures in the inner regions of AGNs with various luminosities.
However, this model is not in contradiction with the obscuring torus paradigm
since the accelerating bi-conical wind could be itself considered as a form of
dusty torus.

In the near-IR, the first $J$, $H$ and $K$-band Adaptive Optics (AO) images of
NGC 1068, at 0.12" resolution (1" corresponds to 72 pc), were reported by Rouan
et al. (\cite{Rouan1998}). They gave an upper limit on the size of the
unresolved K-band core of $\sim$ 9 pc, and highlighted an elongated $S$-shaped
structure radially extending up to 20 pc, at a position angle (PA) $\sim$
102$°$.  Spectroscopic high angular resolution $K$-band observations obtained
with PUEO-GriF at CFHT are presented by Gratadour et al.
(\cite{Gratadour2003}). Emission of hot dust seems to be confined in an almost
resolved region of 120 mas.  Based on $K$, $L$, and $M$-band
diffraction-limited images of NGC~1068 obtained with NACO at the VLT, Rouan et
al. (\cite{Rouan2004}) reported several emissions around the nucleus being
structured in several regions. The $K$ and $L$-band emissions are actually
resolved with a FWHM of 67 and 122 mas respectively, and are elongated in the NS
direction.

First 76 mas resolution $K$-band images of NGC~1068 obtained
 from bispectrum speckle interferometry were interpreted in terms of a resolved
 gaussian component of FWHM $\sim$~30~mas (Wittkowski et al.
 \cite{Wittkowski1998}). Speckle imaging performed by Weinberger et al.
 (\cite{Weinberger}) at the Keck Observatory led to agreeing results.

Then, the first $K$-band long baseline interferometric measurement of NGC~1068
was obtained with VINCI at the VLTI (Wittkowski et al. \cite{Wittkowski2004}). 
Coupled with $K$-band speckle interferometry measurements (Weigelt et al. 
\cite{Weigelt}), they account for the observations with a two-component model
where the small component has an angular size of less than 5 mas and the other
one has a size of 40 mas.  Weigelt et al.  (\cite{Weigelt}) present near-IR
bispectrum speckle interferometry made with the SAO 6m telescope. The $K'$-band
FWHM diameter of the observed compact core is $\sim$ 18$\times$39 mas, and the
PA of the north-western elongation is -16$°$. In the $H$-band, the FWHM
diameter of this same component is $\sim$ 18$\times$45 mas, and the PA is
$\sim$ -18$°$.

First mid-IR images with 0.1" spatial resolution are reported by Bock et al.
(\cite{Bock}), using the 10~m Keck II telescope. They interpret the northern
elongation of the central peak as reprocessed radiation from the AGN.  Tomono
et al.  (\cite{Tomono}) also obtained mid-IR images with 0.1" spatial
resolution with MIRTOS on the 8.2~m Subaru Telescope and confirm that the
central region of NGC 1068 is elongated in the NS direction, with a FWHM of
0.29"~$\times$~0.18". They use a modified grey-body radiation, with absorbing
silicate features in order to model the spectral energy distribution (SED) of
the central peak.

Jaffe et al.  (\cite{Jaffe}) report the first mid-IR $N$-band long-baseline
interferometric measurements using MIDI, the VLTI mid-IR beamcombiner.  They
consider a model composed of two ellipsoidal gaussian disks.  They derive sizes
of about 10~$\pm$~2~mas in the NS direction, parallel to the inner radio jet,
and $<$~12~mas in the direction transverse to the jet for an inner hot
component (T~$>$~800K) and sizes of about 30~$\pm$~5~mas~$\times$~49~$\pm$~4~mas
for an outer warm component (T~$\sim$~350~K).

Infrared interferometry clearly has the potential to study the circum-nuclear
environment of AGNs to bring evidences for or against the unified sheme to help
better understand their nature.  In this paper, we report an alternative
analysis of the first $N$-band MIDI observations of the nucleus of NGC~1068. 
First, in Sect.~\ref{Observations}, we give a short review of the observations
and data processing. Sect.~\ref{Model} details the models considered in this
paper. Then, the results are presented in Sect.~\ref{Results}, and further
discussed in Sect.~\ref{Discussion}. Sect.~\ref{Conclusion} gives the general
conclusions of the study.

%__________________________________________________________________ 
\section{Observations and data reduction} \label{Observations}
%________________________________________________________________

The observations were performed in 2003 (June 15th, 16th; November 9th, 10th),
during four nights of Science Demonstration Time with MIDI, the mid-IR $N$-band
(8-13~$\mu$m) instrument. Observations were carried out with two 8.2~m diameter
Unit Telescope (UT) at the European Southern Observatory (ESO) VLTI on Cerro
Paranal.  A 0.6"~$\times$~2" slit was used with a low resolution prism, set in
the direction perpendicular to the slit, providing a spectral resolution of
$\lambda$/$\Delta \lambda \sim$~30.

MIDI has two interferometric outputs which are the sum of the flux of NGC 1068
and the fringe modulation, plus the background. The NGC 1068 flux and the
modulation were measured separately as the photometric outputs were not
operational at the time of the observation. The two interferograms are
complementary with opposite phases. Using this property, the two outputs are
subtracted to disentangle the modulation from the continuous signals. The
amount of continuous flux from NGC 1068 was accurately obtained thanks to the
``chopping" technique: both UTs alternatively point to the sky and to the
source. Subtraction of the sky sequences from the source sequences yields the
source flux. The chopping frequency was 2~Hz during the two nights.
Since no spatial filter was used at that time, the flux from NGC 1068 can be
considered constant. The modulated part of interferograms is then normalized by
the continuous flux. The contrast of individual fringe scans is
obtained with a classical method which has been used with
precursors of VLTI (Coud\'e Du Foresto et al. \cite{Coude}).
Normalized scan sequences are Fourier transformed. The power spectrum is band
pass filtered in each MIDI spectral channel.  The squared fringe contrast is
the integration of the filtered power spectrum from which the photon and
read-out noise bias have been subtracted.  Statistical errors on squared fringe
contrast estimators are set from the dispersion of measurements.

  %__________________________________________________________________
  \subsection{Visibilities calibration}

  \begin{figure*} \centering \includegraphics[width=12cm]{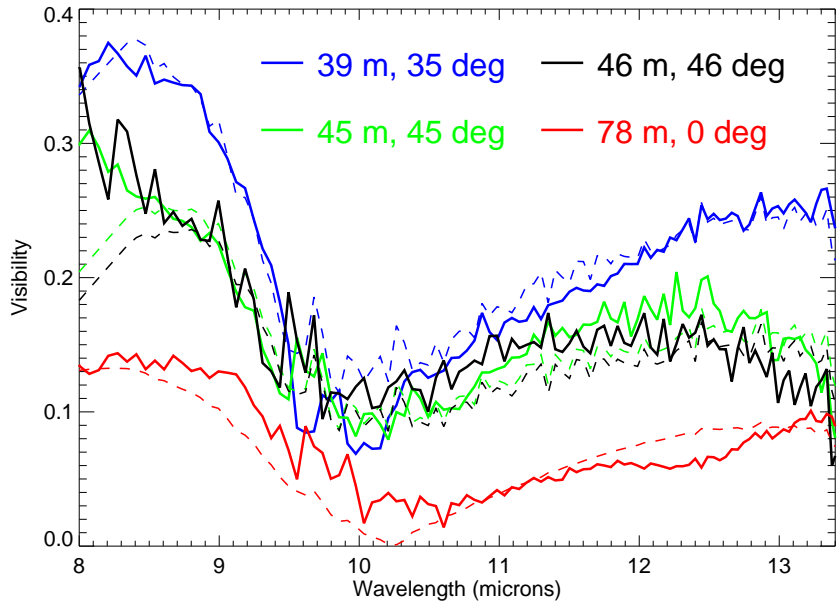}
\caption{Visibilities for the four projected baselines as a function of
wavelengths.  Solid lines correspond to MIDI measurements while dashed lines
are related to fits by the radiative transfer between two spherical components
model (see sect.~\ref{Model} and \ref{Results}).} \label{cb_vis} \end{figure*} 

The entire data reduction has been done with the software developed at LESIA at
Paris Observatory for MIDI\footnote{This software is now available through the
JMMC website: http://mariotti.ujf-grenoble.fr/}. It also relies on
methods developped for precursors of the VLTI (Perrin \cite{Perrin2003}). This software has been used to reduce data on stars
for which visibility functions can be more easily anticipated (Kervella et al. \cite{Kervella}, accepted).

As the pupils of the UTs are larger than the spatial coherence diameter of the
wavefront, and as there was no spatial filtering at the time of the observations,
beams are strongly sensitive to turbulence and alignment defects, causing
coherence losses.  Therefore, visibilities need to be carefully calibrated.  To
do so, a calibration star is observed before and after the observation of the
source. In the case of NGC 1068, the calibrator used was HD 10380 (or $\nu$
Psc), a K3IIIb star whose angular diameter, derived from a photometric scale,
is 2.99 $\pm$ 0.15 mas (Perrin et al. 1998).

The interferometric response of the instrument, the so-called transfer
function, was calculated from the calibrator observations.  The transfer
function is the ratio of the observed fringe contrast by the expected
visibility of the calibrator deduced from the estimated diameter. Plots of the
transfer function as a function of wavelength are also informative about the
quality of the data with respect to turbulent conditions and background
subtraction. Since sensitivity to turbulence decreases with wavelength, the
transfer function increases with wavelength.  Excessive turbulence or
misalignments cause a large gradient between 8 and 13~$\mu$m. Such data are
rejected.

The observations of the source are generally bracketed by observations of the
calibrator star. The source fringe contrasts are obtained by normalizing the
fringe contrast by the transfer function interpolated at the time the source
was observed. A selection is then applied on the histograms of the
squared fringe contrasts of NGC~1068 and of the calibration star. Sequences of
fringe contrasts should have a rough gaussian distribution under steady
turbulence conditions. Sequences of data for which histograms were departing
too clearly from gaussian shape were rejected.

Actually, two calibration stars were observed, $\tau_\mathrm{4}$ Eri and HD
10380. A detailed look at the data led us not to use transfer functions derived
from $\tau_\mathrm{4}$ Eri since they strongly increase up to 11 $\mu$m and
then saturate. As it was the only calibrator star used during the night of
November 10th (except for a single series), and as data obtained on Betelgeuse
were also rejected for this same night (see Perrin et al. \cite{Perrin2005}),
we did not use the data of November 10th for our present study of NGC 1068. In
the same way, we were led to reject the data of June 15th since they most
probably have been affected by large turbulence effects.  A detailed look at
the squared fringe contrast histograms also led us to reject some data from the
two remaining nights. Table~\ref{log_juin} and Table~\ref{log_nov} summarize
the observations eventually kept for the present study.  The combination of
adaptive optics and spatial filtering will eventually solve this issues and
improve the quality of the data.

Visibility measurements of NGC~1068 have been averaged by bins of projected
baselines.  A consistency check has been performed in each bin to assess the
quality of error bars. Ad-hoc error bars have been computed to be consistent
with the dispersion of visibilities when error bars were initially too small.
The four bins are the visibility points used in the following and are presented
in Fig.\ref{cb_vis}. The final absolute accuracy on visibility estimates is on the order
of 3 percent.  Our processing of the data leads to a more extended data set
than that of Jaffe et al.  (\cite{Jaffe}) who kept two visibility points at
baselines of 42 and 72 m.

Two main features are striking on Fig.\ref{cb_vis}.  First, the visibilities
are far smaller than 1, which means that NGC 1068 is well resolved by MIDI. 
Second, visibilities are obviously wavelength dependent, a property which has
to be taken into account in the modeling of the source.

  %__________________________________________________________________
\subsection{Spectral Energy Distribution calibration} \label{SEDcalib}

The Spectral Energy Distribution (SED) as seen by MIDI is of prime importance
to model the spectral and spatial properties of the source. The SED of NGC~1068
is derived from the chopping sequences of the source and of the calibrator. 
However, no observed SED is available for HD~10380 in the
literature. Therefore, we made
use of the calibrated one computed by Cohen et al. (\cite{Cohen}). The integrated flux has been normalized to
that of HD~10380.  The two SEDs are consistent to within 7\%. We eventually
used the mean of these two spectra as a SED for HD~10380. The difference
between the two SEDs was kept as the error on the final SED of HD~10380. A
spectrophotometric response of MIDI was calibrated on HD~10380 using this SED.

Two spectra of NGC 1068 were independently derived from photometries measured
on June 16th and November 9th, 2003. Error bars on photometry measurements are
set from the difference between the spectra measured with the two UTs.  The
final MIDI SED of NGC 1068 is derived from the mean of the spectra of June and
November and from the spectrophotometric response. The difference between the
calibrated SEDs is used as an error estimate.

\begin{figure}[h] \centering \includegraphics[width = 8.5cm]{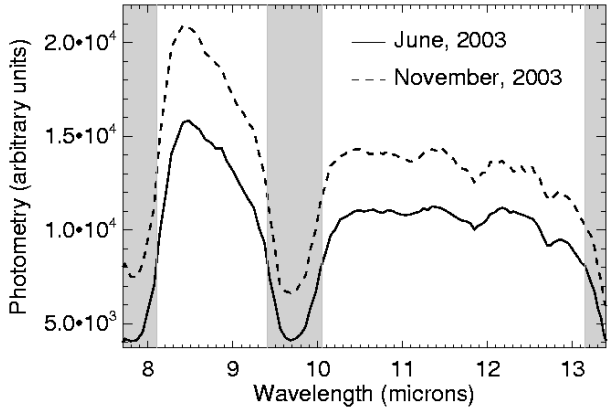}
\caption{NGC~1068 photometry measurements obtained with MIDI on June and
November 2003. Greyed areas are related to wavelength ranges where photometries
are smaller than half of their maximum value.} \label{phot} \end{figure}

\begin{figure*}[t] 
 \centering 
  \includegraphics[width=12cm]{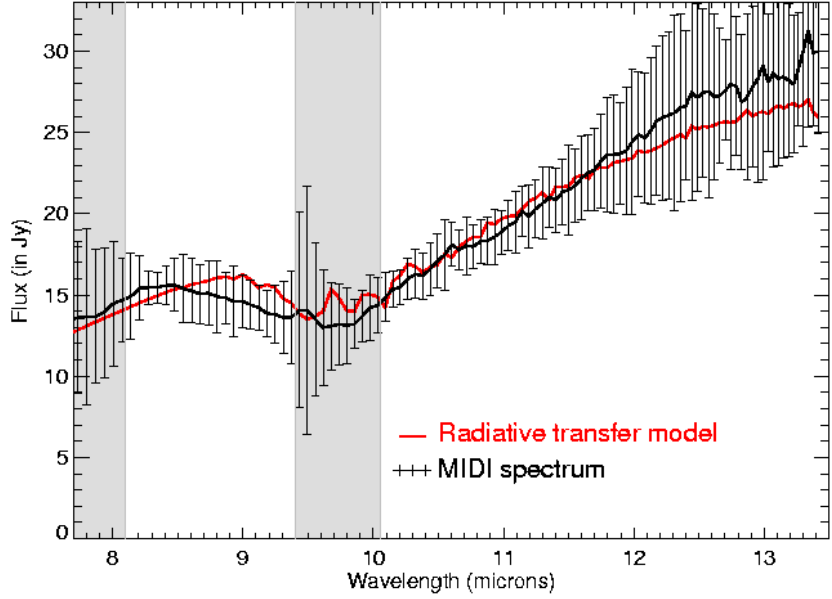}
  \caption{SEDs of NGC~1068 derived from MIDI photometry
    measurements. Black solid line
and vertical error barrs are related to the resulting SED of NGC~1068, which is the
mean between those of June and November.  The
red solid line represent the fit of the SED obtained with the model of
radiative transfer between two spherical components (see sect.~\ref{Model} and
\ref{Results}).  Greyed area correspond to wavelength ranges where photometry
measurements are weak (see fig.~\ref{phot}).} 
  \label{fit_SED} 
\end{figure*}

As shown in Fig.~\ref{phot}, the photometric measurements of NGC~1068 and
HD~10380 are strongly wavelength-dependent. The characteristic loss of signal
around 9.5~$\mu$m is due to absorption by atmospheric ozone.  To avoid too
large noise sensitivity, we do not take into account regions where photometry
measurements are smaller than half the maximum value.  These regions are greyed
on our final spectrum of NGC~1068. The SEDs of June, November and the average
one are presented on Fig.~\ref{fit_SED}.

\begin{figure}[ht] 
 \centering 
  \includegraphics[width = 8.5cm]{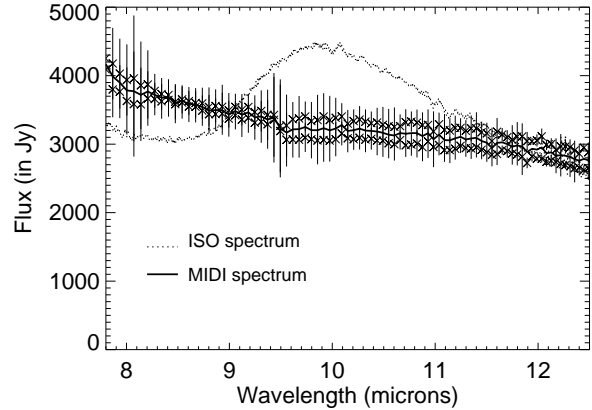} 
  \caption{Comparison between the SEDs of Betelgeuse
derived from MIDI photometry measurements (thick solid line) and the ISO-SWS
spectrum (thin dotted line). This result on
Betelgeuse makes us confident with our calibration procedure and with the
fluxes we derive for NGC~1068 (see text in sect.~\ref{SEDcalib}).} 
 \label{SED_Bet} 
\end{figure}

The SED of NGC~1068 we derived from the MIDI data is different from the one
presented by Jaffe et al. (\cite{Jaffe}): quantitatively, the former is in the
range 13 to 30~Jy whereas the latter is rather between 8 and 15~Jy. Moreover
the slope of the SED we derived after 11~$\mu$m is steeper than the one of Jaffe
et al. Besides, we have validated our calibration procedure on an independent
data set acquired on Betelgeuse. We have checked that we could reproduce the
spectrum of Verhoelst et al. (2005) to an excellent accuracy. Besides, we have
compared our spectrum to the one available from the ISO database (see
Fig.~\ref{SED_Bet}). The field of view of ISO is much larger than that of MIDI
and ISO is detecting the dust shell whose inner edge starts at 0.5'' (Danchi et
al. 1994). The emission feature of silicate dust at 9.7~$\mu$m is therefore
present in the ISO spectrum but not in the MIDI spectrum whose field of view is
reduced to less than 0.3''. Apart from dust, the continua should however
agree.  We find a maximum difference of 16\% between our calibrated SED and the
ISO spectrum outside the silicate dust feature which makes us confident in our
calibration.

%__________________________________________________________________ 
\section{The different steps of data modeling} \label{Model}
%________________________________________________________________

As only a few MIDI data
points on NGC~1068 are available so far, models were computed from a
rigorous stepwise approach on the data rather than from a priori
knowledges about the source. Models remain also simple to avoid
over-interpretations and to keep a reasonnable number of degrees of freedom.

The characteristic apparent size of the object was first
gauged by fitting visibilities with a uniform disk model whose
diameter is wavelength dependent. The 39~m baseline data on the one
hand and the 45-46~m baseline data on the other
hand were fitted independently since these two sets of projected
baselines are separated by 10$°$ in azimuth. The 78~m data were
not used at this stage since they would lay in the second lobe of the uniform disk visibility
function, which model has a poor chance to be the right one at high
spatial frequency. Therefore, these data would
bias the measurement of the characteristic size of the source, which is
the main goal of this first procedure.

\begin{figure}  \centering \includegraphics[height = 6.5cm]{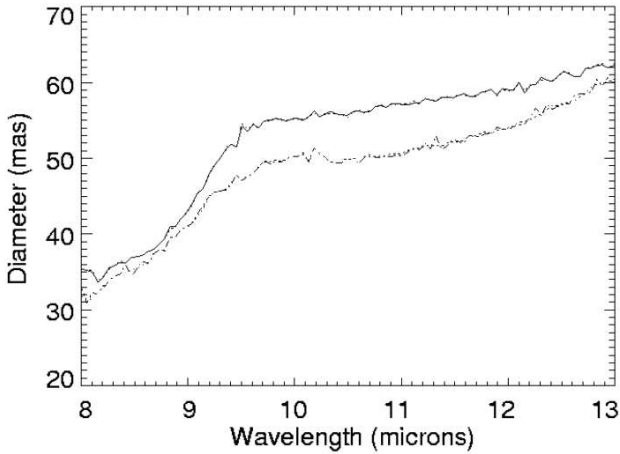}
\caption{Evolution of the diameter derived from the uniform disk model for two
different configurations of the projected baseline (up: 39.59m alone; down:
45.54, 46.63 and 45.7m together).} \label{ponc:fig2} \end{figure}

\begin{figure}  \centering \includegraphics[height = 6.5cm]{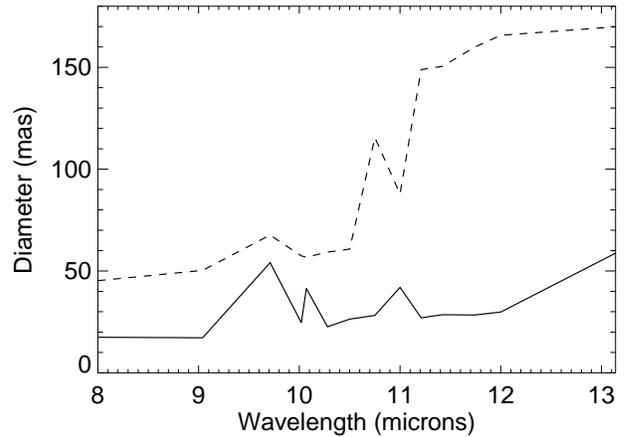}
\caption{Evolution of the diameters of the inner disk (solid line) and the
extended disk (dashed line) derived from the two uniform disks model.}
\label{ponc:fig3} \end{figure}

Fig.~\ref{ponc:fig2} clearly shows the wavelength dependency of the
apparent size of the object. One
may notice that the curves related to the two different baselines do not overlap with a
difference ranging from 2 to 7~mas. First, this may be due to the model
which is clearly too simple. This discrepancy could
then be interpreted as an asymmetry of the nucleus of NGC~1068, which one is well known at larger angular scales.
Thus, to possibly highlight some elongation and orientation of the source, an
attempt was made to fit all the visibility points at the same time
with a uniform elliptical disk. This led to a degeneracy of parameters
which means that extracting informations on the shape and orientation of the
source is not possible with the present MIDI data set. As a
consequence, geometries used in the following modelings were
restricted to spherical symmetry.
The two typical sizes of
$\sim$~30~mas and $\sim$~55-60~mas at 8~$\mu$m and after 9.5~$\mu$m respectively shown up in Fig.~\ref{ponc:fig2} suggest to
elaborate a further model by the addition of a second
uniform disk. This model counts now three free parameters per wavelength, being the diameters of each
disk and the flux ratio between them.  Fits of visibilities are
improved, which strengthens the validity of the two-disk model.
Diameter variations with wavelength are plotted on
Fig.~\ref{ponc:fig3}.

\bigskip

From these previous results, the third step in the modeling is to
consider the radiative transfer between the two spherical components, namely a
compact inner core (corresponding to the inner disk of diameter
$\oslash_\mathrm{in}$) which radiates as a black body with temperature
$T_\mathrm{in}$ and a layer (described by the outer disk of diameter
$\oslash_\mathrm{layer}$) around the compact core which radiates as a thermal
grey-body at $T_\mathrm{layer}$. The absorption in the layer and its emission
are accounted for by its optical depth $\tau (\lambda)$.  No scattering is
considered here.  For the sake of simplicity, we used a model in which the
layer is geometrically thin. This model was successfully applied to Mira stars
and red supergiants in the near-infrared (see e.g.  Perrin et al.
\cite{Perrin2004}).  The interest of this model is that the radiative transfer
can be calculated analytically and that there is no major difference with a
model for which the base of the layer would be at the edge of the compact inner
core.

Thus, if $\oslash_\mathrm{in}$ $\gid$ $\oslash_\mathrm{layer}$ sin$\theta$
(where $\theta$ is the angle between the line of sight and the radius vector of
the two spheres), both contributions of the compact core and of the layer are
observed. Then the emerging intensity can be expressed as:  
  \begin{eqnarray*}
    I(\lambda,\theta)~\sim~B(\lambda,T_{\mathrm{in}})~e^{-~\tau(\lambda)/cos{\theta}}
    \\ +~B(\lambda,T_{\mathrm{layer}})~[1-e^{-~\tau(\lambda)/cos{\theta}}]
  \end{eqnarray*}

In the other case, 
\begin{eqnarray*}
  I(\lambda,\theta)~\sim~B(\lambda,T_{\mathrm{layer}})~[1-e^{-~2\tau(\lambda)/cos{\theta}}]
\end{eqnarray*}

where B($\lambda$,$T$) is the Planck function.

\bigskip 

Thus, as the model was chosen to remain rather simple, the number of
  free parameters was limited to five at each wavelength. Diameters and
  temperatures are global parameters, while the optical depth, chosen to be
  wavelength dependent, allows to account for the visibility variations and
  for the SED. The total number of visibility data is
 $4\times N_\mathrm{\lambda}$ while the number of free parameters is then
  $4+N_\mathrm{\lambda}$ ($N_\mathrm{\lambda}=109$ being the number of spectral
  channels). The diameters are well constrained by the visibilities
  while the latter only provide a one-to-one relation between the
  temperatures. Compatibility with the SED allows to determine a
  unique couple of temperatures.

\bigskip 

The criterion to be minimized is the $\chi^2$, written as: 
\begin{eqnarray*}
\chi^2_\mathrm{Vis} =
\sum_{i=1}^{N}\biggl[\frac{V_i^{2}-M(\oslash_\mathrm{in},
\oslash_\mathrm{layer}, T_\mathrm{in}, T_\mathrm{layer}, \tau_\mathrm{\lambda_i}; S_i)}{\sigma_i}\biggr]^{2} 
\end{eqnarray*}
 
where $M(\oslash_\mathrm{in},\oslash_\mathrm{layer}, T_\mathrm{in},
T_\mathrm{layer},\tau_\mathrm{\lambda} ; S_i)$ is the modeled squared
visibility function, which also depends on the spatial frequency $S_i$ given
by: $S_i = B/\lambda_i$ ($B$ being the projected baseline).

There are several local minima on the $\chi^2$ hyper-surface in parameter
space. Therefore the search for optimal parameter values has to be carried out
in several steps.  To reach the solution diameters, the first step consists in
the exploration of a 2-D grid in
($\oslash_\mathrm{in}$;$\oslash_\mathrm{layer}$) space. For each point on this
grid, the minimization is performed on the temperatures and on a global optical
depth (in other words, a mean optical depth non-dependent on wavelength at this
point of the minimization). For the second step, the same procedure is applied
but on a 2-D grid in ($T_\mathrm{in}$;$T_\mathrm{layer}$) space. The third and
last step consists in running the minimization on optical depth with respect to
wavelength, setting diameters and temperatures at the optimum values found
previously.  Ideally, steps should have been iterated but this did not turn out
to be necessary.

%__________________________________________________________________ 
\section{Results on the NGC 1068 mid-IR core} \label{Results}
%________________________________________________________________

Visibility measurements actually well constrain the geometrical parameters
since the first step of the $\chi^2$ minimization rapidly converges to the
optimum values of geometrical parameters, which are:

  $\oslash_\mathrm{in}$ = 35.3 $\pm$ 3.7~mas 

  $\oslash_\mathrm{layer}$ = 83.1 $\pm$ 6.1~mas 

\bigskip

However, several local minima appear in the $\chi^2$-hypersurface in
temperature space, leading to a degeneracy of these parameters. Indeed,
visibilities are not sensitive to the absolute flux of each component, but to
the relative flux between them.

The degeneracy on temperatures is overcome using the MIDI SED of NGC 1068. The
criterion estimating the goodness of a fit of the SED, with a given couple of
temperatures, is given by: 
\begin{eqnarray*} 
  \chi^2_\mathrm{Spec} = \sum_{i=1}^{N}\biggl[\frac{SED_\mathrm{obs}(i)-SED_\mathrm{fit}(i)}{\sigma_\mathrm{SED}(i)}\biggr]^{2}
\end{eqnarray*}

The optimal couple of temperatures then obtained is:

  $T_\mathrm{in}$ = 361 $\pm$ 12~K 

  $T_\mathrm{layer}$ = 226 $\pm$ 8~K 

\bigskip

  $(\chi^2_\mathrm{Vis})_\mathrm{\textit{Reduced}} $ = 33.9

  $\chi^2_\mathrm{Spec}$ = 16.4

\bigskip

$\chi^2_\mathrm{Vis}$ and $\chi^2_\mathrm{Spec}$ can be minimized independently
as long as minimization of the second one is performed over the local minima of
the first one.

To estimate the confidence interval on parameters, we treat the two couples of
global parameters (i.e.  diameters and temperatures) independently
(e.g. Bevington~\&~Robinson \cite{Bevington}).  For geometrical parameters, we look
at the joint confidence interval given by a variation of 1 with
respect to $\chi^2_\mathrm{Vis}/(\chi^2_\mathrm{Vis})_\mathrm{min}$,
 as a function of $\oslash_\mathrm{in}$ and
$\oslash_\mathrm{layer}$, keeping the other parameters fixed at their optimum
values. This is equivalent to increase the error bars on visibilities
  by $\sqrt{\chi^2/(\chi^2_\mathrm{Vis})_\mathrm{min}}$ and to reaccount
    for the distance to the model. 

For temperatures, the confidence interval corresponds to the
intersection between the $\chi^2_\mathrm{Vis}$ trough in
($T_\mathrm{in}$;$T_\mathrm{layer}$) space and the joint confidence interval on
temperatures which give a reasonable fit of the SED. In order to obtain this
last joint interval, we look at a variation of 1 on the value of
$\chi^2_\mathrm{Spec}/(\chi^2_\mathrm{Spec})_\mathrm{min}$ as a function of $T_\mathrm{in}$ and $T_\mathrm{layer}$.
Values of $\chi^2$ relating to the fits of the visibilities and of the spectrum
are quite high, leading to an under-estimation of the confidence interval on
parameters.  Nevertheless, this must be due on the one hand to the simplicity
of the model considered and on the other hand to the fits of the 45 and 46~m
visbility points between 8 and 8.5~$\mu$m which are poor compared to eslewhere
in the $N$-band (see Fig.~\ref{cb_vis}). Also note on
  Fig.~\ref{cb_vis} that the modeled visibility curves have the same kind of
wavelength dependent fluctuations as the measurements. This is due to the
 third step of the $\chi^2$ minimisation which improves the fit by
 converging to the optimal value of the optical depth at each wavelength.

\begin{figure*}%[t] 
  \begin{center} 
    \includegraphics[height = 6cm]{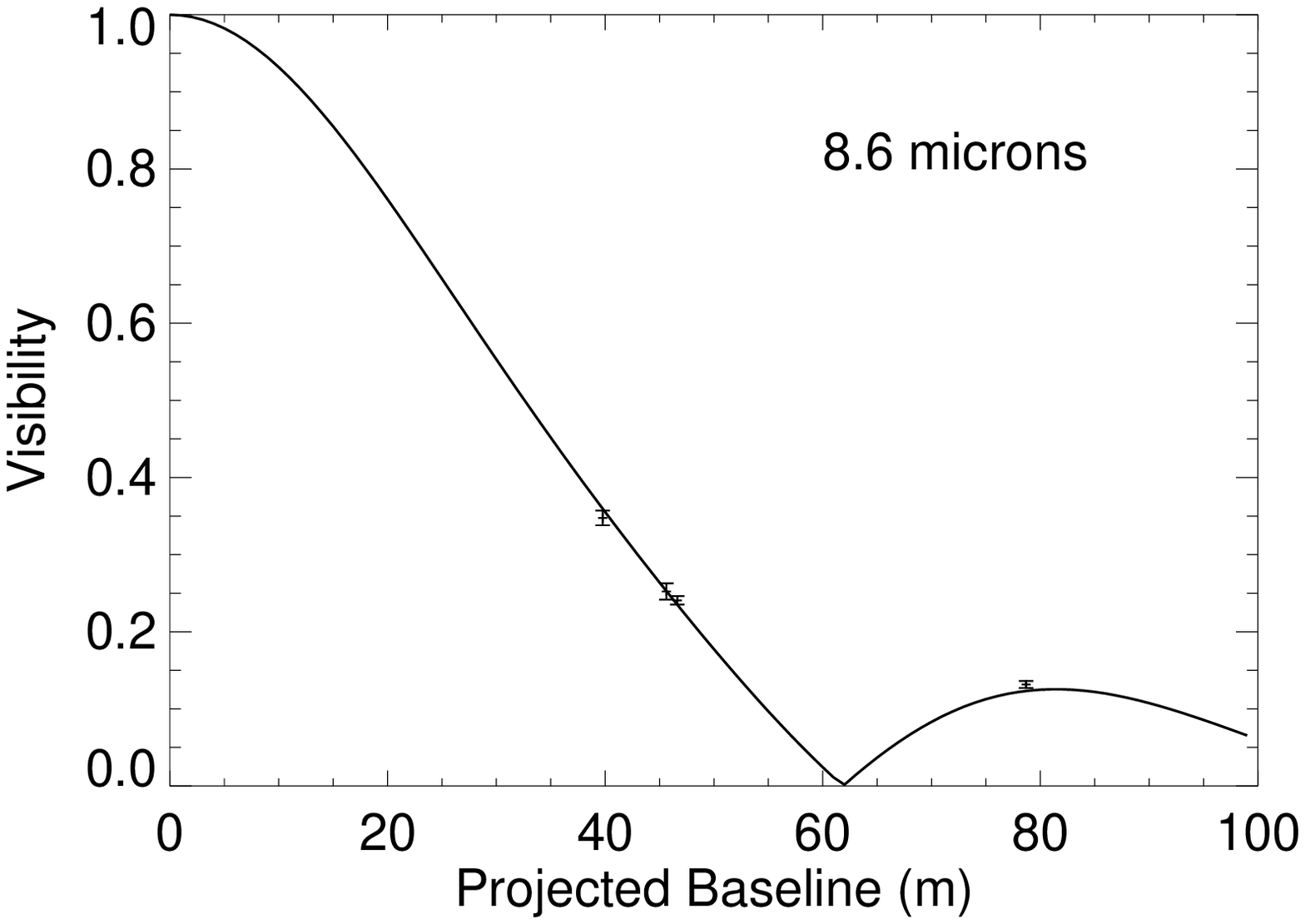} 
    \includegraphics[height = 6cm]{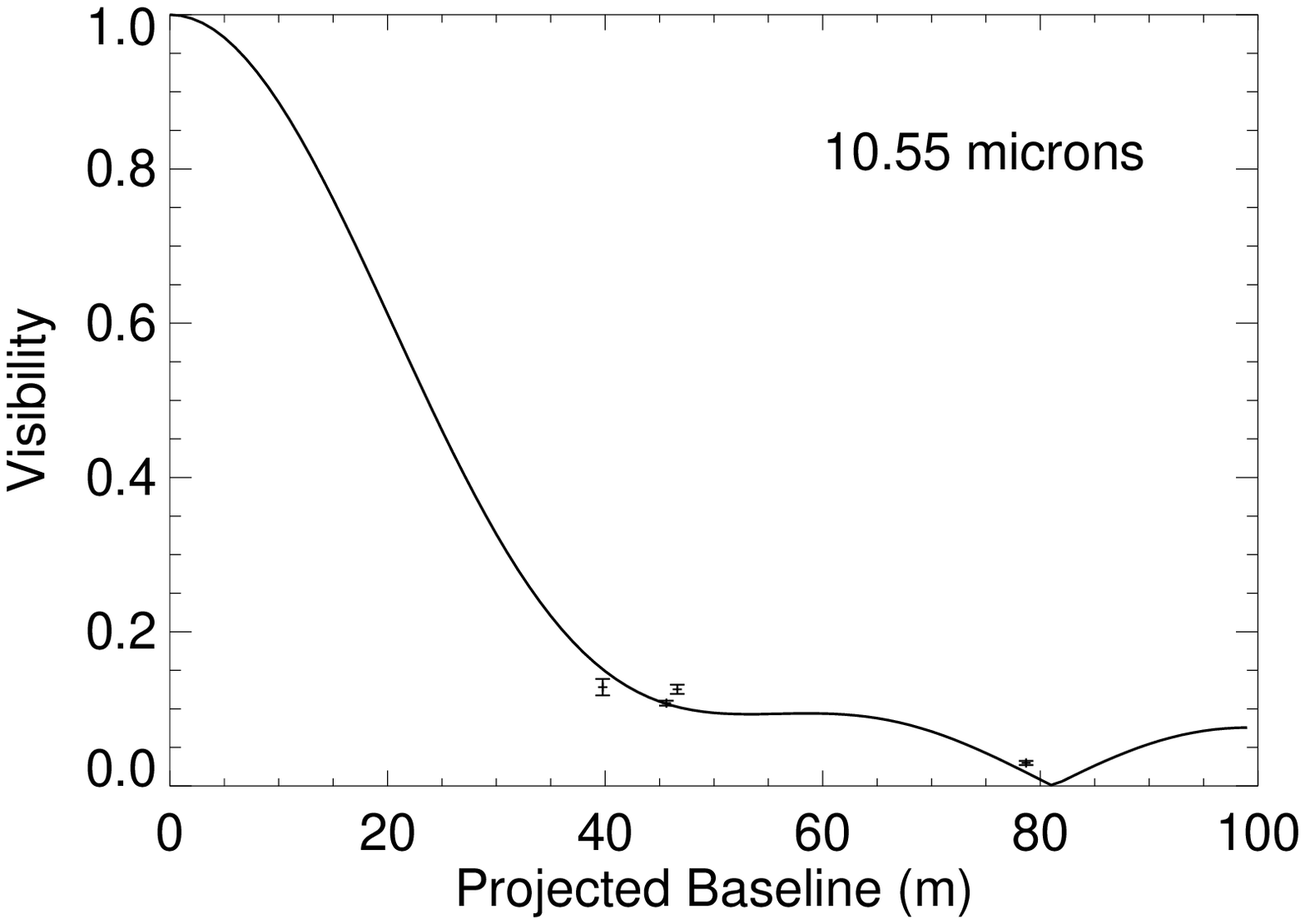}
    \includegraphics[height = 6cm]{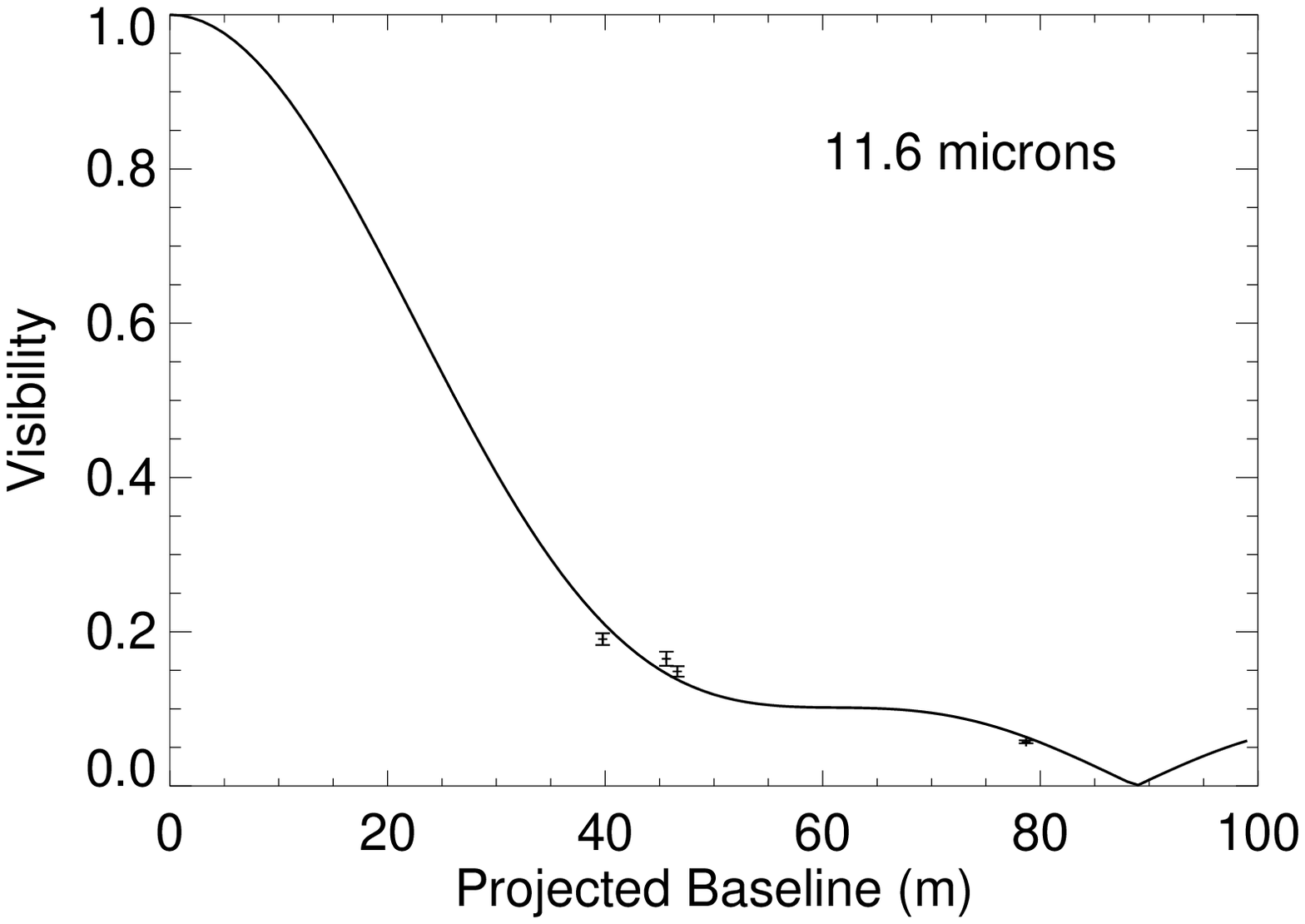} 
    \includegraphics[height = 6cm]{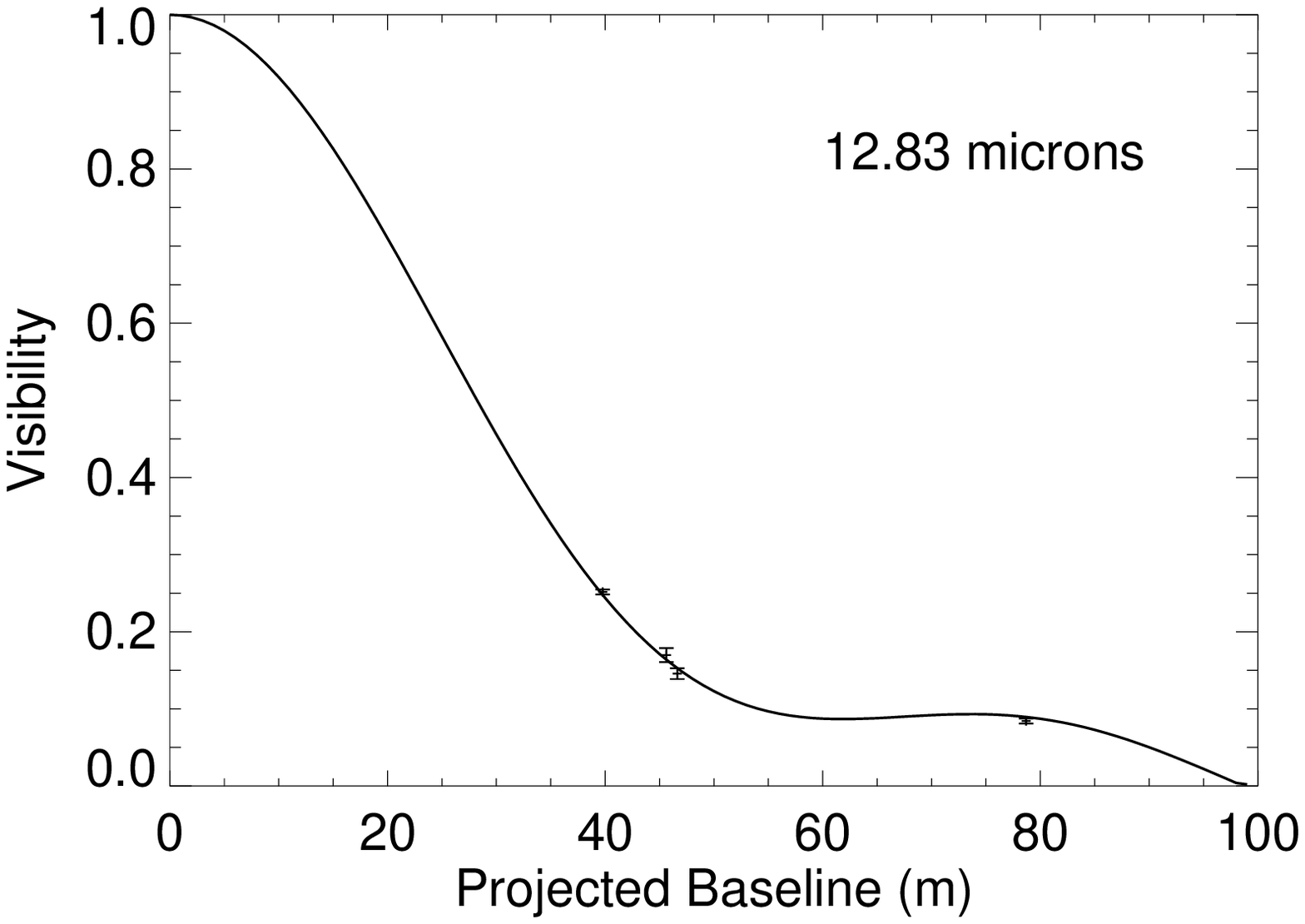} 
    \caption{Fit of the visibilities as a function of the
projected baseline and at given wavelengths with the radiative transfer between
two spherical component model.} 
    \label{fitvisPB} 
  \end{center} 
\end{figure*}

Fig.~\ref{fitvisPB} presents several fits of visibility measurements as a
function of the projected baseline at different fixed wavelengths.  The growing
influence of the extended layer with increasing wavelength is highlighted on
the one hand by the flatness of the visibility function at high spatial
frequencies and on the other hand by the strong slope at low spatial
frequencies. Moreover, the figures underline the embarrassing lack of
visibility points at low projected baselines. As the slope at low spatial
frequencies is really model-dependent, these points are needed to put stronger
constrains on the modeling.

\begin{figure}[t] 
\centering 
\includegraphics[height = 6cm]{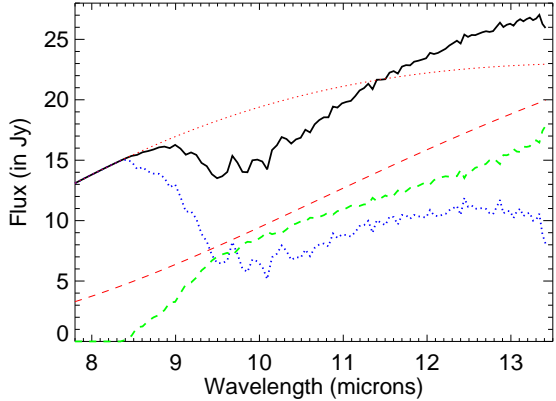}
\caption{Contributions of each spherical components of the radiative transfer
model to the SED. \textit{Red thin dotted line}: flux emitted by the
non-absorbed inner component. \textit{Blue thick dotted line}: fraction of the
flux emitted by the inner component and absorbed through the layer. \textit{Red
thin dashed line}: flux emitted by the layer without its self-absorption
considered. \textit{Green thick dashed line}: flux emitted by the layer and
self-absorbed by it.} \label{Contrib_SED} \end{figure}

The fit of the MIDI spectrum is presented on Fig.~\ref{fit_SED}. This shows
that not only the simple radiative transfer model accounts for MIDI visibility
measurements, but also it gives a good representation of the SED across the
full $N$-band. Fig.~\ref{Contrib_SED} presents the contribution of each
component to the spectrum. We observe that the inner compact component is the
only one to contribute to the SED until 8.5~$\mu$m while the flux emitted by
the layer begins to dominate above 9.4~$\mu$m.  This figure also shows that, in
the framework of our model, the gap of the SED around 10~$\mu$m is not due to
the crossing of the two black body emissions, but indeed to a peak in the
absorption by the layer.

Therefore, the originality of the present study resides in its capability to
provide the evolution of the optical depth of the extended component as a
function of wavelength in the $N$-band, without any a priori assumption on it
like its variation law or the composition of the extended layer.  The optical
depth resulting from the whole analysis is plotted as a function of wavelength
in Fig.~\ref{ponc:fig9} (dark solid line) and will be discussed further in
sect.~\ref{Absorption}.

%__________________________________________________________________
\section{Discussion} \label{Discussion}
%________________________________________________________________

  %__________________________________________________________________
  \subsection{Spatial description of the source} \label{Global_desc}

As the number of visilitity measurements is extremely small and as the
(u,v) plane is not sufficiently covered, models able to fit the MIDI
  data points are degenerated. This lack of sensitivity in the models
  highlights the importance of following a stepwise approach from the data.
  
The MIDI visibility points can be fitted by a different and a priori
model that consists in a two gaussian elliptical disk model (Jaffe et al.
\cite{Jaffe}). The derived sizes are $\sim10\times12$~mas for an inner hot,
optically thick component and $\sim30\times49$~mas for a warm component.
The factor of 2 between these values and the ones we derived
with our model is actually explained by the differences
between the respective geometries considered.

Comparing mid- and near-IR observations, it appears that the largest
$K$-band structure corresponds in size to the smallest structure
  detected in the
mid-IR with MIDI. From $K$-band bispectrum speckle interferometry, Wittkowski
et al. (\cite{Wittkowski1998}) interpreted the observed central core of NGC~1068 as a slightly resolved gaussian disk of FWHM $\sim$~30~mas. Speckle
imaging from the Keck Observatory also showed that the flux emitted by the core
of the AGN may be accounted for by an unresolved point source whose size is
smaller than $\sim$~20~mas and which is surrounded by an emitting region
extending up to 10~pc (Weinberger et al. \cite{Weinberger}).  Wittkowski et al.
(\cite{Wittkowski2004}) combined near-IR speckle data with the $K$-band
visibility point obtained with VINCI and interpreted the whole dataset with a
two-component model: a small one having a size of $\sim$~5~mas (or 0.4~pc)
responsible for 40\% of the $K$-band flux and an extended one of $\sim$~40 mas.
In addition, near-IR bispectrum speckle interferometric observations led to a
N-W extended structure in $K'$-band having a size of $39\times18$~mas and a
P.A.  $\sim$~-16$°$ (Weigelt et al.  \cite{Weigelt}). With regard to the AO
observations performed with NACO at VLT, the $K$-band core is slightly N-S
elongated with a FWHM of 67~mas (Rouan et al., \cite{Rouan2004}). The spatial
description of the source derived from speckle observations is also fully
consistent with the best resolution AO observations performed with NAOS-CONICA
which show a resolved structure in the $Ks$-band, elongated along the same P.A.
$\sim$~-16$°$ and of FWHM $\sim30\times<15$~mas after deconvolution
($80\times65$~mas before deconvolution; Gratadour et al.
\cite{Gratadour2005}). The sizes derived from $K$-band and $N$-band
observations underline the consistency between the near-IR and the mid-IR
observations of the nucleus of NGC~1068.

According to Maloney (\cite{Maloney}), the dusty torus has suitable
temperatures (T~$\sim$~250 K) and H$_2$ molecules densities
($n_\mathrm{H_\mathrm{2}}\sim10^{8}-10^{10}$~cm$^{-3}$), in order
to produce the $6_{16}\rightarrow5_{23}$ rotational transition at 22~GHz
of the H$_2$O masers (e.g., Elitzur \cite{Elitzur}, chap.  10).  H$_2$O masers
can be produced by the absorption by a H$_2$O molecule of an X-ray emitted by
the central engine of the AGN, and the 22~GHz radio emission is expected to
trace the inner edge of the torus. Sub-milliarcsecond angular resolution
observations made on NGC~1068 with the Very Long Baseline Interferometer (VLBI)
showed that the redshifted emission is divided in four groups, linearly
distributed between 6 and 14~mas from the core along a direction at 45$°$ from
the radio synchrotron jet axis (Greenhill et al. \cite{Greenhill}).  Additional
observations of the blueshifted emission showed that masers are actually
spatially distributed on a thin ring inclined at -45$°$, on scales extending
from 0.65 to 1 pc (i.e., 9.3 and 15.7 mas; Greenhill \& Gwinn \cite{G&G}). 
Thus, the ring traced by the H$_2$O masers seems to correspond to the inner
edge of the dusty torus (Kartje et al. \cite{Kartje}). This is actually what we
are observing since the masers are located at the same place as the $K$-band
structure derived from the VINCI/Speckle observations and also observed with
NAOS-CONICA, and just inside the extended component derived from our modeling
of the MIDI data.

Moreover, as claimed by Rouan et al. (\cite{Rouan2004}) and Gratadour et al. 
(\cite{Gratadour2005}), and according to current estimates of the dust
sublimation radius, graphite grains are able to survive at such small distances
from the core of the nucleus of NGC 1068, so that this dust species could be
part of the observed $K$-band structure. Indeed, a theoretical estimate of the
dust sublimation radius is given by:

\begin{equation}  
r_\mathrm{sub} =
1.3~~(L_\mathrm{UV,46})^{1/2}~(T_\mathrm{1500})^{-2.8} ~~$pc$ 
\end{equation} 
where $L_\mathrm{UV,46}$ is
the UV luminosity (in units of 10$^{46}$~erg.s$^{-1}$) and $T_\mathrm{1500}$ is
the sublimation temperature of the dust species considered (in units of 1500 K;
Barvainis \cite{Barvainis}). Upper limits are then obtained by setting
the UV luminosity of the central source to the observed bolometric luminosity
of the nucleus and by considering sublimation temperatures of 1400~K and 1750~K
for silicates and graphite grains respectively. Pier et al. (\cite{Pier1994})
investigated the bolometric luminosity of the nucleus of NGC~1068. Depending on
the fraction of nuclear flux reflected into our line of sight which has been
determined by several different ways, the bolometric luminosity of the nucleus
of NGC~1068 ranges from $\sim6.8\times10^{43}$~erg.s$^{-1}$ to
$\sim3.4\times10^{45}$~erg.s$^{-1}$, the most probable value being on
the order of $\sim3.4\times10^{44}$~erg.s$^{-1}$ (Pier et al.
\cite{Pier1994}). These dispersed values lead to a sublimation radius ranging
from $\sim$~0.07 to 0.50~pc (i.e. $\sim$~1 to 7~mas) for graphite grains and
from $\sim$~0.13 to 0.91~pc (i.e. $\sim$~1.8 to 12.6~mas) for silicate
grains, the most probable
values being $\sim$~0.16~pc and $\sim$~0.3~pc (or $\sim$~2.2~mas and $\sim$~4.2~mas) respectively.

As a conclusion, in spite of the great uncertainties on the
  estimation of sublimation radii, it is remarkable
that H$_2$O masers seem to be confined in an area located between the
sublimation radius of silicate grains and the outer edge of the inner component
we have inferred from mid-IR data. In addition, according to near-IR
observations of the nucleus of NGC~1068, it seems that the dust surrounding the
central engine is divided in several layers of which the first one -- emitting
in $K$-band and probably extending from the sublimation radius of graphite
grains up to about $\sim$~1.2~pc -- could be mainly composed of graphite grains
and would harbour the observed H$_2$O masers.

  %__________________________________________________________________
  \subsection{Temperature of the MIDI compact component}

The temperatures we derive are much cooler than those of Jaffe et al.
(\cite{Jaffe}). There are different reasons to these discrepancies. The
most important one comes from the difference in the geometries considered that
leads to a factor of 2 between the respective sizes derived and consequently to a factor
of almost 0.5 on the temperatures. Moreover, our spectral calibration yields
twice as much flux. Finally, contrary to Jaffe et al., we do not consider
absorption in the inner compact component which we modeled as a pure black
body. Indeed, for Jaffe et al., the optical depth associated to silicates in
the inner hot component is about 2.1, meaning that the radiation emitted by the
central engine is greatly attenuated in this optically thick medium. In this
paper, we do not consider this additive parameter and the opacity of the inner
component affects the final temperature.

The temperatures we derive from the radiative transfer modeling seem very low
with respect to dust sublimation temperatures commonly assumed for graphite
grains and silicates which are on the order of 1700 and 1450~K respectively.
However, following the approach of Krolik (\cite{Krolik}) who, for first rough
estimates, assumes thermal emissions by spherical black bodies, the typical
angular size of each component inside an AGN is given by the expression:

\begin{equation} \theta = 87 ~\Biggl(\frac{F_\mathrm{obs}}{10^{-11}
erg.cm^{-2}.s^{-1}}\Biggr)^{1/2}
~\Biggl(\frac{T_\mathrm{eff}}{10^{3}K}\Biggr)^{-2} ~\mu \mathrm{as}
\end{equation}

Hence, for typical sizes of 35 and 83~mas and a flux of $\sim$~17~Jy at
8~$\mu$m and $\sim$~30~Jy at 13~$\mu$m (according to the MIDI spectrum of NGC~1068), the typical temperatures obtained with this formula are $\sim$~260 and
160~K. These values are consistent with those we obtain with the radiative
transfer model. This shows that considering this simple description, even at
small distance from the central engine, temperature falls quickly. Thus, the
apparent weakness of temperatures derived from our modeling is finally not
surprising.

Furthermore, according to SED measurements of the central elliptical region of
$290\times180$~mas of the nucleus of NGC~1068 carried out at the Subaru
telescopes with a 0.1" spatial resolution, Tomono et al. (\cite{Tomono})
derived a temperature of $\sim$~234~K and an optical depth
$\tau_{9.7\mu m}\sim$~0.92, from a model of modified greybody radiation absorbed by
silicates. Temperature for such a large area is consistent with those we obtain
and the optical depth we are able to derive at 9.7~$\mu$m is fairly close.

  %__________________________________________________________________
\subsection{Dust absorption in the layer} \label{Absorption}

The optical depth as a function of wavelength across the $N$-band is
deduced from the optimisation of the fit of visibility measurements. Therefore,
it is rather model-dependent.

\begin{figure} \centering \includegraphics[width = 8.5cm]{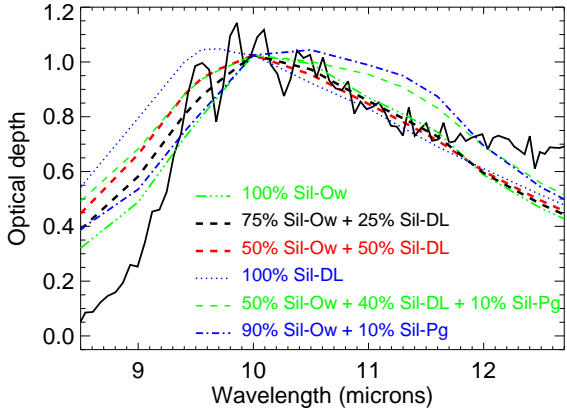}
\caption{Optical depth as a function of the wavelength derived from the
radiative transfer model (thick line); fits provided by the DUSTY code with
different compositions of silicates are superposed. Sil-Ow refers to warm
siliactes from Ossenkopf (\cite{Ossenkopf}); Sil-DL refers to silicates from
Drain \& Lee (\cite{D&L}); SiC-Pg refers to $\alpha$ silicon carbide from
P\'egouri\'e (\cite{Pegourie}).} \label{ponc:fig9} \end{figure}

However, a detailed study of the shape of the optical depth obtained
with our model is interesting. Indeed, looking at  Fig.~\ref{ponc:fig9}, it
seems that the strong slope between 8.5 and 9.5~$\mu$m and the bump around 10~$\mu$m can be the signature of the presence of amorphous silicates in the
composition of the layer (in comparison with the absorption features at 10~$\mu$m associated with the Sgr A* region observed with ISO for instance; Kemper
et al. \cite{Kemper}). Moreover, given the sublimation radius of silicates
which is on the order of 0.64~pc (or $\sim$~8.9~mas; see
sect.~\ref{Global_desc}), this type of dust is actually able to survive in the
layer.  Therefore, to verify the probable presence of silicates, we tried to
fit the shape of the derived optical depth with DUSTY, a radiative transfer
code into a dusty layer, which includes absorption, emission and scattering by
dust (Ivez\`ic, Nenkova \& Elitzur, \cite{Ivezic}). This code handles spherical
geometry and has been used by Imanishi \& Ueno (\cite{Imanishi2000}) to model
the Cygnus~A nucleus absorption feature at 9.7~$\mu$m. It is important to
underline here that if the torus is viewed almost edge-on (the common
assumption for NGC~1068), the use of spherical symmetry in the modeling does
not affect the derived optical depth since the thickness of material viewed
along the line of sight remains the same as in the case of a torus-like
geometry.

Next, Maiolino et al. (\cite{MaiolinoI}, \cite{MaiolinoII}) claim that dust in
the circumnuclear region of AGNs must have different properties than in the
galactic interstellar medium (ISM) and that silicate grain sizes must be larger
than 3~$\mu$m in order to explain the lack of silicate features in ISO spectra
of Seyfert II nuclei. This is confirmed by the computation of the
$\tau_{3.4}/\tau_{9.7}$ ratio which is $\sim$~0.11 for NGC~1068 (Dartois et al.
\cite{Dartois}).  Indeed, this ratio would be $\sim$~0.06-0.07 assuming the
$A_V/\tau_{9.7}$ ratio of both the galactic diffuse interstellar medium (DISM)
in the local 3~kpc around the Sun (Roche \& Aitken \cite{R&A1984}) and in the
direction of the Galactic Centre (Roche \& Aitken \cite{R&A1985}), and the
$A_V/\tau_{3.4}$ ratio in these different directions (Pendleton et al. 
\cite{Pendleton}). In addition to grain sizes, such a difference in extinction
laws of the Galactic ISM and Seyfert~II AGNs can also result from a temperature
gradient in the obscuring dust (Imanishi \cite{Imanishi}).  Evidences of the
flattening of the silicates optical depth with the growth of grain sizes are
presented by Bouwman et al. (\cite{Bouwman}) and Kemper et al. (\cite{Kemper}).

\begin{figure} \centering \includegraphics[width = 8.5cm]{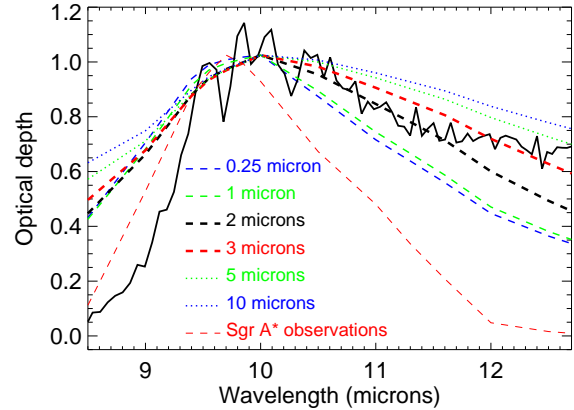}
\includegraphics[width = 8.5cm]{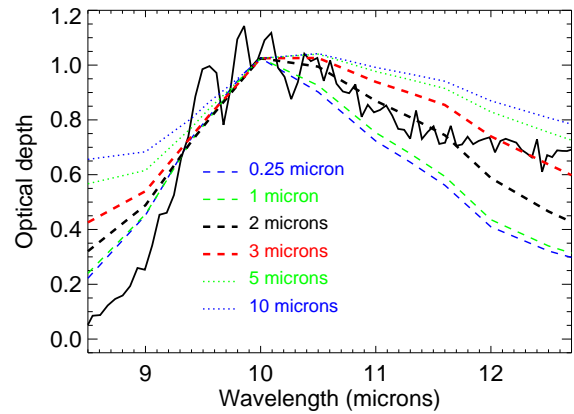} \caption{Fit of the optical
depth provided by the DUSTY code for various maximum sizes of grains, for a MRN
(Mathis, Rumpl \& Nordsieck \cite{MRN}) grain size distribution and two
different compositions. \textit{Top:} 50$\%$ silicates from Ossenkopf
(\cite{Ossenkopf}) and 50$\%$ silicates from Drain \& Lee (\cite{D&L}), and
superposition with silicate features intrinsic to the Sgr A* region.
\textit{Down:} 100$\%$ silicates from Ossenkopf (\cite{Ossenkopf}).}
\label{ponc:fig10} \end{figure}

Following these considerations, tentative fits of the derived optical depth
with several simple compositions of amorphous silicates and various grain sizes
provided with the code DUSTY are presented in Fig.~\ref{ponc:fig9}
and~\ref{ponc:fig10}. We are considering the MRN distribution (Mathis, Rumpl \&
Nordsieck \cite{MRN}) which consists in a power law (index~=~-3.5) distribution
of grain sizes ranging from a minimum value of 0.05~micron to various maximum
sizes. Species of dust available in DUSTY are ``astronomical silicates" and
graphite grains from Draine \& Lee (\cite{D&L}; referred as Sil-DL and Grf-DL
thereafter), warm oxygen-deficient and cool oxygen-rich silicates --
representative of circumstellar and interstellar silicates -- from Ossenkopf
(\cite{Ossenkopf}; referred as Sil-Ow and Sil-Oc), $\alpha$ Silicon carbid from
P\'egouri\'e (\cite{Pegourie}; referred as SiC-Pg), and amorphous carbon
(referred as amC).  A first remark is that compositions of Sil-Ow or Sil-Oc
alone will provide the same optical depth (in shape and amount); the addition
of grf-DL or amC in the composition of the dust has no effect on the shape of
the optical depth law since absorption or emission features of these species
appear around 4~$\mu$m. We therefore only considered Sil-Ow, Sil-DL and SiC-Pg
compounds thereafter.

Sil-DL alone provides a good fit of the optical depth after 10~$\mu$m while the
10~$\mu$m silicate feature is globally shifted towards lower wavelengths (see
Fig~\ref{ponc:fig9}). On the contrary, when SiC-Pg are added, the bump is
shifted to the right and it does not account for the shape around 11~$\mu$m. 
Best fits are obtained with a mixture of Sil-Ow with a small amount of Sil-DL.

Next, for the study of grain sizes, we considered two of the well fitting
compositions: 50\% Sil-DL + 50\% Sil-Ow and 100\% Sil-Ow. We first considered a
single grain size distribution. As shown by Kemper at al. (\cite{Kemper}), we
observed that the silicate feature is significantly shifted towards longer
wavelengths as the grain size increases. Best fits are provided by a size of
1~micron. Nevertheless, the MRN distribution provides better fits.  According
to Fig.~\ref{ponc:fig10} which presents comparisons of fits provided by various
maximum grain sizes (ranging from 0.25 to 10~microns), distributions of
relatively small grains (with sizes smaller than 3~microns) account better for
the bump around 10~$\mu$m while larger grains flatten the optical depth law
beyond 10~$\mu$m.

The quality of the fit of the optical depth is poorer at the edges of
the band, and the optical depth derived from the modeling appears to be more
asymmetric than what would be expected from DUSTY. There are several
reasons for that: first,
the model considered here for NGC~1068 is a fairly simple
one. Consequently it is not
meant to be a real description of the source compared to its a
priori complexity. Thus, differences between the real geometry and physical properties of
the source and the model will result in inaccuracies in the estimate of the
optical depth. Second, differences may arise from the physical properties
which are modeled. No scattering is taken into account in the two component
radiative transfer model used to fit the MIDI data. Moreover, following Kemper
et al. (\cite{Kemper}) who assume that the spectra of optical depth are all the
narrower as emission by silicates increases, the narrowness of the shape of the
optical depth provided by DUSTY beyond 11.5~$\mu$m could highlight that our
model under-estimate silicate emissions in the layer.

However, if we now compare the evolution of optical depth with the one provided
by the ISO observations of Sgr~A$*$ (Kemper et al. \cite{Kemper}; see
Fig.~\ref{ponc:fig10}), it appears that, although the silicates peak is
displaced with respect to wavelength, both plots have similar slopes at low
wavelengths. Besides, despite there are no evidences yet of the presence of PAH
(Polycyclic Aromatic Hydrocarbon molecules) features in the ISO spectrum of NGC~1068 (at 7.7, 8.6 or 11.3~$\mu$m; Sturm et al.  \cite{Sturm}), this trail
should be investigated in a further study.  Similarly, the strong pure-rotation
lines of H$_2$O could significantly contribute to opacity across the $N$-band
and could also provide an explanation for the wing in the upper half of the
band. Therefore, as these molecules would be able to survive in an environment
such as the dusty layer, their presence could also affect the shape of the
optical depth and be responsible for the non perfect fit.

Nevertheless, at this point conclusions can already be drawn from the present
study of the optical depth.  The main features of the dust around the core of
NGC 1068 could be: \begin{itemize} \item a mixture of amorphous silicates from
Draine and Lee (\cite{D&L}) with a large fraction of cosmic silicates from
Ossenkopf (\cite{Ossenkopf}). \item small grains, with sizes ranging from 0.05
to 3 microns. \end{itemize} The properties derived from the optical depth
deduced from the MIDI observations show that dust around the central engine of
the nucleus of NGC 1068 has been detected.  This is the first direct evidence
of the presence of amorphous silicates in the composition of the dusty
environment at radial distances of 1 to 2 pc from the central engine.

  %__________________________________________________________________
  \subsection{Proposed description of the dust distribution}

\begin{figure} \centering \includegraphics[width = 8.5cm]{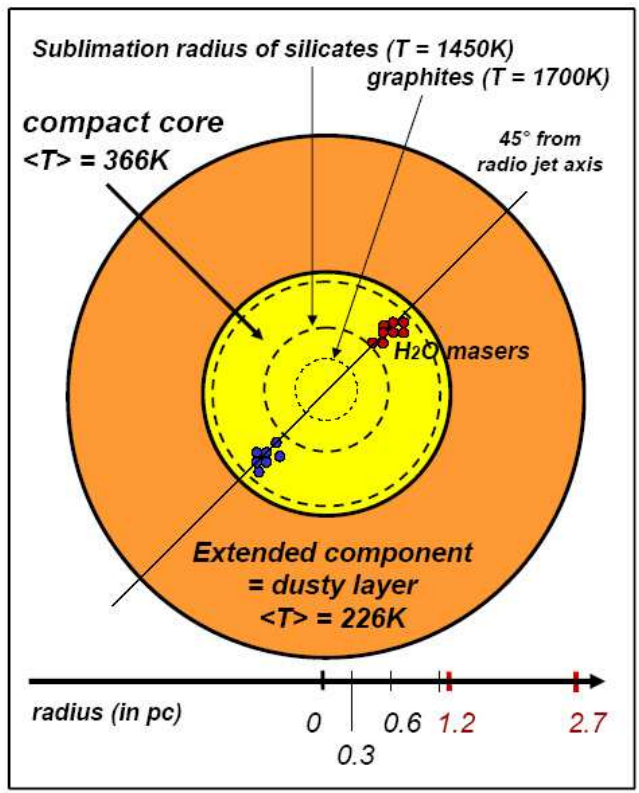}
\caption{Comparison between the results derived from the radiative transfer
model between two spherical components applied to MIDI data with others
observations.} \label{ponc:fig14} \end{figure} 

From the above discussion, a picture of the distribution of dust inside the
core of NGC 1068 can be drawn (see the scheme in Fig.~\ref{ponc:fig14}). It is
a layered structured. One of the layers is detected in the $K$- and $N$-bands
and extends from the sublimation radius of graphite grains up to about 1.2 pc. 
It could be made of graphite grains, and silicates in its outer parts. The
second one emitting in the $N$-band and extending from 1.2 pc up to about 2.7
pc, is well resolved by MIDI and seems to be warmer (about 200 K) and made of
amorphous silicates.  This picture is consistent with H$_2$O maser emission
since masers need a dusty environment in order to survive in the hard UV field
radiated by the central engine. The first layer would then act as a protecting
shield and would be a favorable environment for water maser emissions. It also
provides an explanation for the low temperatures derived from the radiative
transfer model applied to the MIDI observations.  Indeed, there would be a
strong radial temperature gradient inside the inner layer.

%______________________________________________________________ 
\section{Conclusions} \label{Conclusion}
%________________________________________________________________

We present an independent analysis of the first VLTI MIDI observations of the
nucleus of NGC~1068 obtained in 2003. The small visibilities measured across
the $N$-band show that the core of NGC~1068 is well resolved in the mid-IR.

After a severe selection of the data, different basic models with an increasing
number of free parameters have been tried to fit both interferometric and
spectral data.  As a final scenario, we adopt a simple radiative transfer model
with two concentric spherical components. This model accounts well for the MIDI
visibilities and SED across the $N$-band. The inner compact component has a
radius of about 17~$\pm$~2~mas and a temperature of about 361~$\pm$~12~K, and the outer layer
extends up to 41~$\pm$~3~mas from the core of the nucleus and has an average
temperature of about 226~$\pm$~8~K. Results are consistent with observations in other
spectral bands of the thermal IR and with H$_2$O masers emissions. One
specificity of this approach is that it provides the evolution of the layer
optical depth across the N band, without requiring assumptions about its
nature. Although this parameter is rather model-dependent, its shape
seems to be characteristic of amorphous silicate grains. Thus, the conclusion
to be drawn from the analysis of the data with this simple model is that MIDI
has actually observed the dust distribution around the core of the nucleus of
NGC 1068 emitting in the $N$-band. 

Besides, this study highlights the degeneracy in the numbers of models
being able to describe the current MIDI observations. This should be overcome
thanks to various coming near- and mid-IR observations at the VLT. Concerning
MIDI, the coverage of the (u,v) plane needs to be increased in order to get
visibility points for several new lengths and orientations of the
projected baseline. Observations with MIDI should besides
put additional constrains on the modeling and lead to a convergent
description of the morphology and of the physical properties of the
dust distribution. 
Next, as the slope of the visibility function
considered in the model presented by Jaffe et al. (\cite{Jaffe})
differs from ours at low spatial frequency, visibility points at low projected baseline
should lead to the choice of the most favorable model. As the shortest projected
baseline reachable with the VLTI and the UTs is $\sim$~30~m, one would have to
make use of VISIR (the VLT imager and spectrometer for the mid-IR) to
get 8.2~m baseline visibility points.

This study is a very first step in the understanding of the dusty
environment of the core of an AGN. It will be mandatory to test the approach
presented here with observations of other AGNs such as Centaurus A or
Circinus.

\begin{acknowledgements}  We thank the MIDI and the Science Demonstration teams
for providing the data, Emmanuel Dartois for enlightening information on grain
physics, Damien Gratadour and Daniel Rouan for discussions, Tijl Verhoelst for
providing synthetic and ISO spectra for Betelgeuse, Arcturus and
HD~10380, Walter Jaffe for discussions about the Leiden data reduction.
\end{acknowledgements}

%______________________________________________________________  
% Bibliography:

\clearpage
\onecolumn

%__________________________________________________ One column table
   \begin{table}
      \caption{Log of June 16, 2003 observation, performed between UT1 and UT3} % titre du tableau
      \label{log_juin} % pour referencer le tableau dans le texte
      \centering % pour centrer le tableau
      \begin{tabular}{c c c c c c c}
        \hline\hline
	UT time & Projected baseline & Baseline Azimuth & UT time & UT
	time \\
        & (m) & ($°$) & (Calibrator 1) & (Calibrator 2)\\
	\hline
09:56:54.000 & 78.6670 & 0.036 & 09:34:22.000 & - \\  
09:58:46.811 & 78.6670 & 0.036 & 09:35:38.451 & - \\
10:02:32.432 & 78.6670 & 0.036 & 09:36:54.901 & - \\
10:10:12.000 & 78.6670 & 0.036 & \\
10:11:44.611 & 78.7050 & 0.036 & \\
10:13:17.223 & 78.7050 & 0.036 & \\
10:14:49.834 & 78.7050 & 0.036 & \\
	\hline
	\end{tabular}
	\end{table}

%________________________________________________ One column table
   \begin{table}
      \caption{Log of November 9, 2003 observations, performed
      between UT2 and UT3} % titre du tableau
      \label{log_nov} % pour referencer le tableau dans le texte
      \centering % pour centrer le tableau
      \begin{tabular}{c c c c c c c}
        \hline\hline
	UT time & Projected baseline & Baseline Azimuth & UT time & UT
	time \\
        & (m) & ($°$) & (Calibrator 1) & (Calibrator 2)\\
	\hline

02:50:22.451 & 39.5900 & 35.16 & - & 03:59:13.432 \\  
02:52:55.352 & 39.5900 & 35.16 & - & 04:00:29.882 \\  
& & & - & 04:01:46.333 \\
\\
04:48:28.000 & 45.5410 & 44.60 & 03:59:13.432 & 05:17:54.000 \\  
04:49:44.451 & 45.5410 & 44.60 & & 04:00:29.882 & 05:19:10.450 \\  
04:51:00.901 & 45.5410 & 44.60 & & 04:01:46.333 & 05:20:26.901 \\  
\\ 
05:58:27.451 & 46.6340 & 45.90 & 05:17:54.000 & 06:33:28.000 \\
05:59:43.882 & 46.6340 & 45.90 & 05:19:10.450 & 06:34:44.450 \\
& & & 05:20:26.901 & 06:36:00.901 \\
\\ 
07:01:56.000 & 45.7090 & 44.76 & 06:33:28.000 & - \\
07:03:12.450 & 45.7090 & 44.76 & 06:34:44.450& - \\
 & & & 06:36:00.901 & - \\  

  \hline
\end{tabular}
\end{table}


\begin{thebibliography}{}

  \bibitem[1985]{Antonucci} Antonucci, R.R.J., \& Miller, J.S. 1985, ApJ, 297, 621

  \bibitem[1987]{Barvainis} Barvainis, R. 1987, ApJ, 320, 537

  \bibitem[2004]{Beckert} Beckert, T., \& Duschl, W. J. 2004, A\&A, 426, 445

  \bibitem[1992]{Bevington} Bevington, P.R., \& Robinson D.K. 1992, McGraw-Hill Editions

  \bibitem[2000]{Bock} Bock, J.J., Neugebauer, G., Matthews, K. et
  al. 2000, ApJ, 120, 2904

  \bibitem[2001]{Bouwman} Bouwman, J., Meeus, G., de Koter, A. 2001,
  A\&A, 375, 950

  \bibitem[1999]{Cohen} Cohen, M., Walker, R.G., Carter, B., et al. 1999, AJ, 117, 1864

  \bibitem[1997]{Coude} Coud\'e Du Foresto, V., Ridgway, S.  T.,
  Mariotti, J. M. 1997, A\&AS, 121, 379

  \bibitem[Danchi et al. 1994]{danchi1994} Danchi, W.C., Bester, M.,
  Degiacomi, C.G., Greenhill, L.G., Townes, C.H. 1994, AJ, 107, 1469

  \bibitem[2004]{Dartois} Dartois, E., Marco, O., Munoz-Caro, G. M. et
  al. 2004, A\&A, 423, 549

  \bibitem[1984]{D&L} Draine, B.T. \& Lee, H.M. 1984, ApJ, 285, 89

  \bibitem[1995]{Efstathiou} Efstathiou, A., Hough, J.H., \& Young,
  S. 1995, MNRAS, 277, 1134

  \bibitem[1992]{Elitzur} Elitzur, M.  1992, Astronomical Masers,
  Kluwer Academic Publishers

  \bibitem[2000]{Elvis} Elvis, M. 2000, ApJ, 545, 63

  \bibitem[2003]{Galliano} Galliano, E., Alloin, D., Granato, G. L. et
  al. 2003, A\&A, 412, 615

  \bibitem[1993]{Granato} Granato, G.L., \& Danese, L.  1993, MNRAS, 268, 235

  \bibitem[1996]{Greenhill} Greenhill, L.J., Gwinn, C.R., Antonucci,
  et al. 1996, ApJ, 472, L21

  \bibitem[1997]{G&G} Greenhill, L.J., Gwinn, C.R. 1997, Ap\&SS, 248, 261G

  \bibitem[2003]{Gratadour2003} Gratadour, D., Cl\'enet, Y., Rouan,
  D., et al. 2003, A\&A, 411, 335

  \bibitem[2005]{Gratadour2005} Gratadour, D., Rouan, D., Mugnier,
  L. M., et al. 2005, A\&A, submitted

  \bibitem[2000]{Imanishi} Imanishi, M. 2000, MNRAS, 319, 331

  \bibitem[2000]{Imanishi2000} Imanishi, M.  \& Ueno, S.  2000, ApJ, 535, 626

  \bibitem[1999]{Ivezic} Ivez\`ic, Z., Nenkova, M.  \& Elitzur, M.,
  1999, User Manual for DUSTY, University of Kentucky Internal Report

  \bibitem[2004]{Jaffe} Jaffe, W., Meisenheimer, K., R\"ottgering,
  H.J.A., et al. 2004, Nature, 429, 47

  \bibitem[1999]{Kartje} Kartje, J.F., K\"onigl, A., \& Elitzur,
  M. 1999, ApJ, 513, 180

  \bibitem[2004]{Kemper} Kemper, F., Vriend, W.  J., \& Tielens,
  G. G. M. 2004, ApJ, 609, 826

  \bibitem[2005]{Kervella} Kervella, P., et al. accepted for publication in A\&A

  \bibitem[1999]{Krolik} Krolik, J., H., 1999, Active Galactic Nuclei,
  Princeton Series in Astrophysics

  \bibitem[1988]{K&B} Krolik, J. H., \& Begelman, M. 1988, ApJ, 329, 702

  \bibitem[2001a]{MaiolinoI} Maiolino, R., Marconi, A., Salvati, M.,
  et al. 2001, A\&A, 365, 28

  \bibitem[2001b]{MaiolinoII} Maiolino, R., Marconi, A., \& Oliva,
  E. 2001, A\&A, 365, 37

  \bibitem[2002]{Maloney} Maloney, P.R. 2002, PASA, 19, 401M

  \bibitem[1977]{MRN} Mathis, J.  S., Rumpl, W., \& Nordsieck, K. H. 1977, ApJ, 217, 425

  \bibitem[1991]{Miller} Miller, J.S., Goodrich, R.W., Mathews,
  W.G. 1991, ApJ, 378, 47

  \bibitem[2002]{Nenkova} Nenkova, M., Ivezic, Z., \& Elitzur,
  M. 2002, ApJ, 570, L9

  \bibitem[1992]{Ossenkopf} Ossenkopf, V., Henning, Th. \& Mathis,
  J.S. 1992, A\&A, 261, 567

  \bibitem[1994]{Pendleton} Pendleton, Y.  J., Sandford, S.  A.,
  Allamandola, L. J., et al. 1994, ApJ, 437, 683

  \bibitem[1988]{Pegourie} P\'egouri\'e, B. 1988, A\&A, 194, 335

  \bibitem[Perrin et al. (1998)]{perrin1998} Perrin, G., Coud\'e du
  Foresto, V., Ridgway, S.T., et al. 1998, A\&A, 331, 619

  \bibitem[2003]{Perrin2003} Perrin, G. 2003, A\&A, 398, 385

  \bibitem[2004]{Perrin2004} Perrin, G., Ridgway, S.T., Mennesson,
  B. et al. 2004, A\&A, 426, 279

  \bibitem[2006]{Perrin2005} Perrin, G., et al. A\&A, in preparation

  \bibitem[1992]{PierKrolik1} Pier, E. A., \& Krolik, J. H. 1992, ApJ, 401, 99

  \bibitem[1993]{PierKrolik2} Pier, E. A., \& Krolik, J. H. 1993, ApJ, 418, 673

  \bibitem[1994]{Pier1994} Pier, E. A., Antonucci, R., Hurt, T. et
  al. 1994, ApJ, 428, 124

  \bibitem[1984]{R&A1984} Roche, P. F., \& Aitken, D. K. 1984, MNRAS, 208, 481

  \bibitem[1985]{R&A1985} Roche, P. F., \& Aitken, D. K. 1985, MNRAS, 215, 425

  \bibitem[1998]{Rouan1998} Rouan, D., Rigault, F., Alloin, D., et
  al. 1998, A\&A, 339, 687

  \bibitem[2004]{Rouan2004} Rouan, D., Lacombe, F., Gendron, E., et
  al. 2004, A\&A, 417, L1

 \bibitem[2005]{Schartmann} Schartmann, M., Meisenheimer, K.,
 Camenzind, M. et al. 2005, A\&A, 437, 861

  \bibitem[2000]{Sturm} Sturm, E., Lutz, D., Tran, D., et al. 2000,
  A\&A, 358, 481

  \bibitem[1997]{Thatte} Thatte, N., Quirrenbach, A., Genzel, R., et
  al. 1997, ApJ, 490, 238

  \bibitem[2001]{Tomono} Tomono, D., Doi, Y., Usuda, T.  et al. 2004,
  A\&A, 417, 1

  \bibitem[Verhoelst et al. 2005]{verhoelst2005} Verhoelst, T., Decin,
  L., Van Malderen, R., et al. 2005, submitted to A\&A

  \bibitem[2004]{Weigelt} Weigelt, G., Wittkowski, M., Balega, Y.Y.,
  et al.  2004, A\&A, 425, 77

  \bibitem[1999]{Weinberger} Weinberger, A. J., Neugebauer, G., and
  Matthews, K. 1999, ApJ, 117, 2748

  \bibitem[1998]{Wittkowski1998} Wittkowski, M., Balega, Y., Beckert,
  T., et al. 1998, A\&A, 329, L45

  \bibitem[2004]{Wittkowski2004} Wittkowski, M., Kervella, P.,
  Arsenault, R., et al. 2004, A\&A, 418, L39

  \bibitem[2002]{Zier} Zier, C., \& Biermann, P. L. 2002, A\&A, 396, 91

\end{thebibliography}
\end{document}